\documentclass[superscriptaddress,twocolumn]{revtex4-2}

\usepackage{graphicx}
\usepackage{dcolumn}
\usepackage{bm}
\usepackage[version=4]{mhchem}
\usepackage{amsmath}%

\begin{document}

\title{Deep-level structure of the spin-active recombination center in dilute nitrides}

\author{A. C. Ulibarri}
\email{agathaulibarri@yahoo.com}
\affiliation{Laboratoire de physique de la matière condensée, CNRS, Ecole Polytechnique, IP Paris, 91128 Palaiseau, France}
 
\author{C. T. K Lew}
\affiliation{Centre of Excellence for Quantum Computation and Communication Technology, School of Physics, University of Melbourne,
Melbourne, VIC 3010, Australia}

\author{S. Q. Lim}
\affiliation{Centre of Excellence for Quantum Computation and Communication Technology, School of Physics, University of Melbourne,
Melbourne, VIC 3010, Australia}

\author{J. C. McCallum}
\affiliation{School of Physics, University of Melbourne, Parkville, VIC, Australia}

\author{B. C. Johnson}
\affiliation{School of Science, RMIT University, Melbourne 3001, Australia}

\author{J. C. Harmand}
\affiliation{Centre de Nanosciences et de Nanotechnologies, CNRS, Université Paris-Saclay, 91120 Palaiseau, France}

\author{J. Peretti}
 \affiliation{Laboratoire de physique de la matière condensée, CNRS, Ecole Polytechnique, IP Paris, 91128 Palaiseau, France}

 \author{A. C. H. Rowe}
 \email{alistair.rowe@polytechnique.edu}
 \affiliation{Laboratoire de physique de la matière condensée, CNRS, Ecole Polytechnique, IP Paris, 91128 Palaiseau, France}

\date{\today}

\begin{abstract}
A Gallium interstitial defect (Ga$_{\textrm{i}}$) is thought to be responsible for the spectacular spin-dependent recombination (SDR) in GaAs$_{1-x}$N$_x$ dilute nitride semiconductors. Current understanding associates this defect with two in-gap levels corresponding to the (+/0) and (++/+) charge-state transitions. Using a spin-sensitive photo-induced current transient spectroscopy, the in-gap electronic structure of a $x$ = 0.021 alloy is revealed. The (+/0) state lies $\approx$ 0.27 eV below the conduction band edge, and an anomalous, negative activation energy reveals the presence of not one but \textit{two} other states in the gap. The observations are consistent with a (++/+) state $\approx$ 0.19 eV above the valence band edge, and a hitherto ignored, (+++/++) state $\approx$ 25 meV above the valence band edge. These observations can inform efforts to better model the SDR and the Ga$_{\textrm{i}}$ defect’s local chemical environment. 

\end{abstract}

\maketitle

Spin-dependent Shockley-Read-Hall recombination (SDR) at a paramagnetic recombination center couples the minority carrier charge dynamics to their spin via a dynamic polarization of the centers \cite{weisbuch1974}. In order to observe SDR, conduction electrons must be spin-polarized, and this is most commonly achieved using large, static magnetic fields. On resonance in a radio-frequency field, the SDR is then revealed in a measurable quantity related to the charge dynamics, for example the photo-luminescence (PL) intensity \cite{geschwind1959} or the photo-current (PC) \cite{lepine1972}. This is the basis for optically- and electrically-detected magnetic resonance (ODMR and EDMR respectively), methods that have become important in the context of spin-based quantum technologies \cite{awschalom2018}.

When the band-to-band optical selection rules permit the optical orientation of non-equilibrium conduction electron spins \cite{zakarchenya1984}, it is also possible to observe SDR in zero field \cite{weisbuch1974, kalevich2005}. The effect is particularly striking in dilute nitrides of the form GaAs$_{1-x}$N$_x$ where increases in PL intensities up to one order of magnitude are observed when passing from a linearly- to a circularly-polarized pump \cite{kalevich2005, kalevich2007}. ODMR indicates that the paramagnetic center responsible for the SDR in these materials is the Gallium interstitial, Ga$_\textrm{i}$, in the (++) charge state \cite{wang2009,wang2009b}, although the exact details of the local alloy disorder is unclear \cite{laukkanen2012}. While there is therefore partial information available on the crystallographic nature of the spin-active defect, nothing is known about its electronic structure despite the fact that this is fundamental to its characteristics as a mediator of extremely rapid SDR \cite{kalevich2007}. This absence is addressed here using a novel, light-polarization-dependent form of photo-induced current transient spectroscopy \cite{balland1986} (or pol-PICTS) that provides an alternative means to achieve spin sensitivity in a deep-level transient spectroscopy \cite{myers2022}.

\begin{figure}
\includegraphics[width=1\columnwidth]{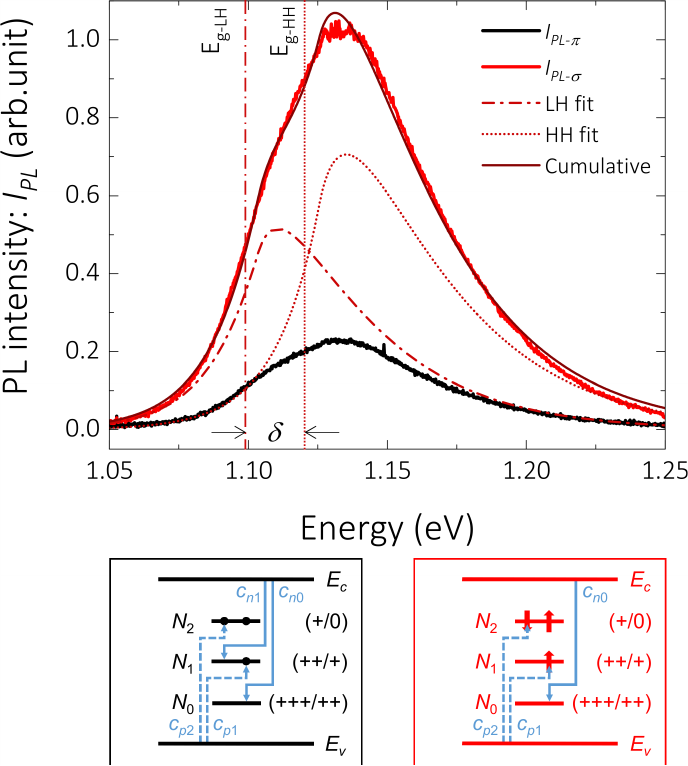}
\caption{\label{PLspectra} Room-temperature PL spectra of GaAs$_{0.979}$N$_{0.021}$ excited with a $\pi$- (black) and a $\sigma$- (red) polarized pump. The large PL intensity increase under a $\sigma$-pump is due to SDR \cite{kalevich2005}. The insets show the allowed capture processes (electrons with solid lines, holes with dotted lines) under the two pump polarizations. The minimum energy gap determined from the spectrum is $\textrm{E}_{\textrm{g-LH}} = $ 1.09 eV.}
\end{figure}

The alloy studied here is a p-type ($p$ = 10$^{18}$ cm$^{-3}$), 50 nm thick epilayer of GaAs$_{1-x}$N$_x$ grown by molecular beam epitaxy onto a GaAs substrate \cite{harmand2002}. When continuously photo-excited at a wavelength of 887 nm, PL spectra of the form shown in Fig. \ref{PLspectra} are obtained. The SDR is apparent from the factor of 5 increase in the PL intensity when switching from a linearly-polarized pump ($\pi$, black spectrum) to a circularly-polarized pump ($\sigma$, red spectrum). The insets provide a schematic explanation of how SDR arises. The three states of the centre corresponding to the (+/0), (++/+), and (+++/++) charge state transitions are shown lying in the gap between the conduction band edge, $E_c$, and the valence band edge, $E_v$ \cite{laukkanen2012}. In what follows the names and densities of these three states are labelled using a subscript corresponding to the electron occupation numbers, $N_2$, $N_1$, and $N_0$ respectively \cite{ivchenko2010}. 

With a $\pi$-polarized pump the conduction electrons and valence holes are unpolarized and four capture processes are possible as shown in Fig. \ref{PLspectra} (black inset); electron capture on $N_1$ at rate $c_{n1}$ to form $N_2$, electron capture on $N_0$ at rate $c_{n0}$ to form $N_1$, hole capture to $N_1$ at rate $c_{p1}$ to form $N_0$, and hole capture on $N_2$ at rate $c_{p2}$ to form $N_1$. If the $\sigma$-polarized pump results in a 100 \% spin polarized conduction electron population, the paramagnetic $N_1$ states become 100 \% polarized dynamically \cite{weisbuch1974,kalevich2005,ivchenko2010}. The exchange interaction on the centre then forbids electron capture to $N_1$ \cite{kaplan1978} as indicated in Fig. \ref{PLspectra} (red inset). If conduction electrons are not fully polarized this process is merely suppressed. Its suppression or absence increases the conduction electron lifetime yielding the observed PL intensity increase. For the SDR to be large as observed, either $c_{n1} \gg c_{n0}$ or the $N_0$ state should be absent. The first condition is related to a difference in capture cross sections as discussed below. However, it is the latter assumption which is generally made in the literature \cite{ivchenko2010}.

Of importance here is the estimation of the semiconducting gap from the shape of the spectrum as described in detail elsewhere \cite{ulibarri2023}. A two Roosbroeck-Shockley component fit to the red spectrum in Fig. \ref{PLspectra} reveals that the heavy holes (red, dotted fit) and light holes (red, dash-dot fit) have different gaps labeled $\textrm{E}_{\textrm{g-LH}} = $ 1.09 eV and $\textrm{E}_{\textrm{g-HH}} =$ 1.12 eV respectively. The splitting between the two, $\delta =$ 30 meV, corresponds to an alloy containing 2.1 \% nitrogen i.e. $x$ = 0.021. The minimum gap of the GaAs$_{0.979}$N$_{0.021}$ is thus $\textrm{E}_{\textrm{g-LH}}$ which will be used below.

The GaAs$_{0.979}$N$_{0.021}$ is electrically contacted with two micro-bonded aluminum wires separated by approximately 170 $\mu$m  which facilitates measurement of a photo-current when the 887 nm pump is focused to a $\approx$ 6 $\mu$m spot between them, and when a voltage of -7 V is applied. The SDR displays a characteristic peaked power dependence \cite{ivchenko2010}, and it is found that maximum SDR is achieved here for a 35 mW pump. This pump power and applied voltage is used throughout (see Supplementary Material).

\begin{figure}
\includegraphics[width=0.96\columnwidth]{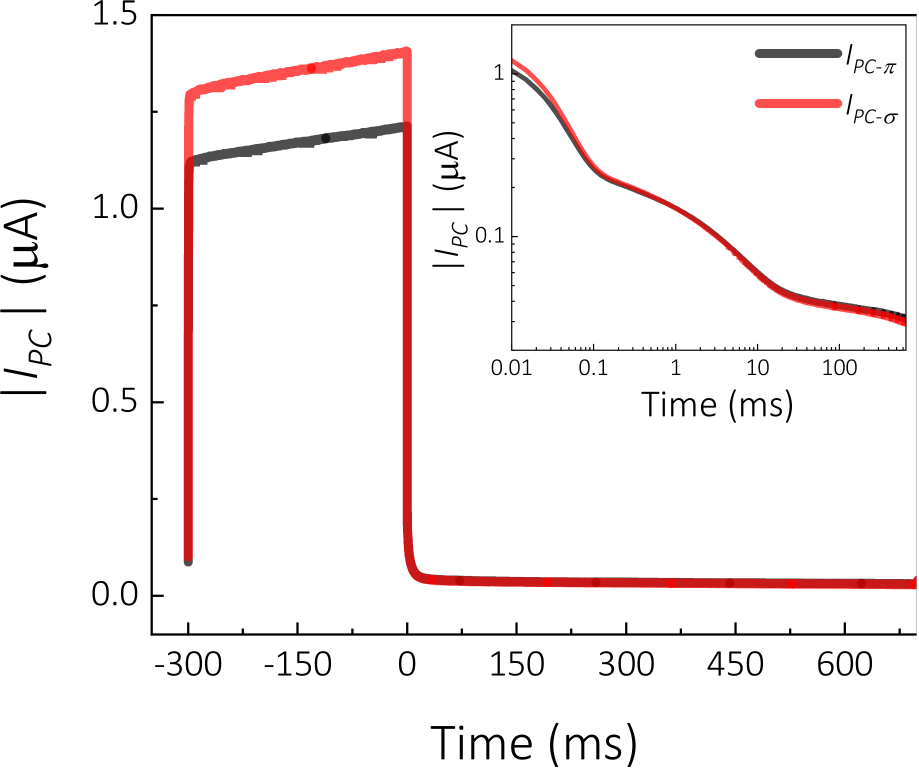}
\caption{\label{transients} Typical PICTS sequence with either a $\sigma$-polarized (red) or a $\pi$-polarized (black) filling pulse. The SDR appears as an increase in the PC measured under a $\sigma$-polarized pump. The PC transients measured over 700 ms after the filling pulse shows non-exponential behavior (inset).}
\end{figure}

In a PICTS measurement, the photo-excitation amplitude is modulated in time. Here the so-called filling pulse during which the sample is optically pumped is 300 ms long and the following dark period, achieved using a fast electro-optic modulator as a switch, is 700 ms long. It is during this dark period that the PC transient is measured at a 200 kHz sampling rate. A typical time trace of the photo current is shown in Fig. \ref{transients}.

During the filling pulse a larger absolute PC, $\| I_{PC} \|$, is measured when using a $\sigma$-pump due to SDR. The increase is smaller than that recorded in the PL spectra of Fig. \ref{PLspectra} because of contacting and transport effects \cite{kunold2011}. Importantly, the PC transient during the return-to-equilibrium following the filling pulse is not a single exponential as seen in the log-log plot shown inset in Fig. \ref{transients}. This situation if frequently encountered in real semiconductors and requires analysis with either traditional boxcar methods \cite{balland1986} or the more modern inverse Laplace transform approach \cite{evans2000}. 

\begin{figure}
\includegraphics[width=1\columnwidth]{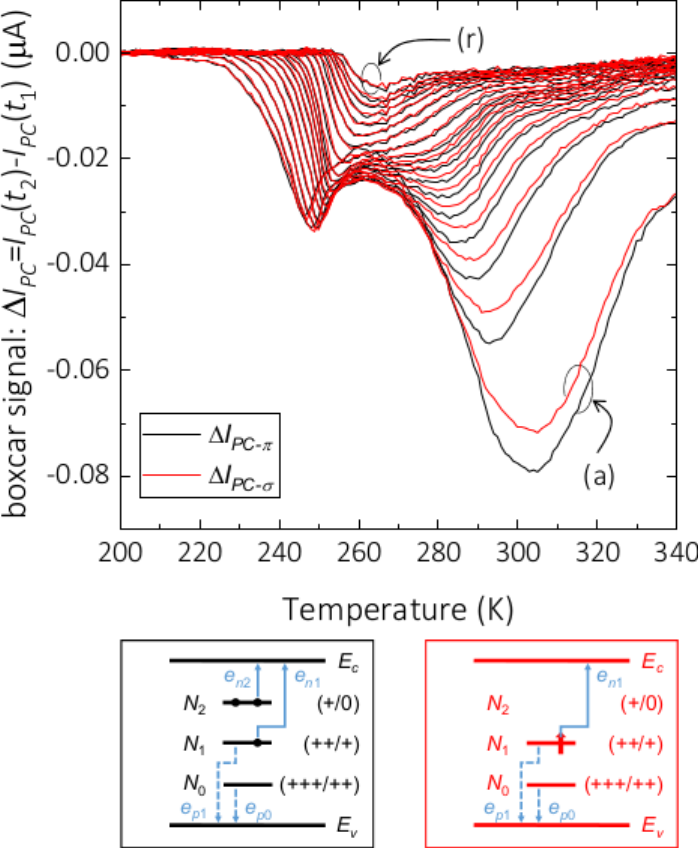}
\caption{\label{boxcar} The boxcar signal obtained with a $\pi$- (black) and $\sigma$- (red) polarized filling pulses. Two peaks are clear to the eye. The amplitude of the higher temperature peak is polarization-dependent and is associated with electron emission from the $N_2$ state (see text). The lower temperature of the two is independent of polarization and exhibits an anomalous shift to higher temperatures as the rate window is shifted to lower rates. The insets show the allowed emission processes following $\pi$- (black) and $\sigma$- (red) polarized filling pulses.}
\end{figure}

\begin{figure}
\includegraphics[width=1\columnwidth]{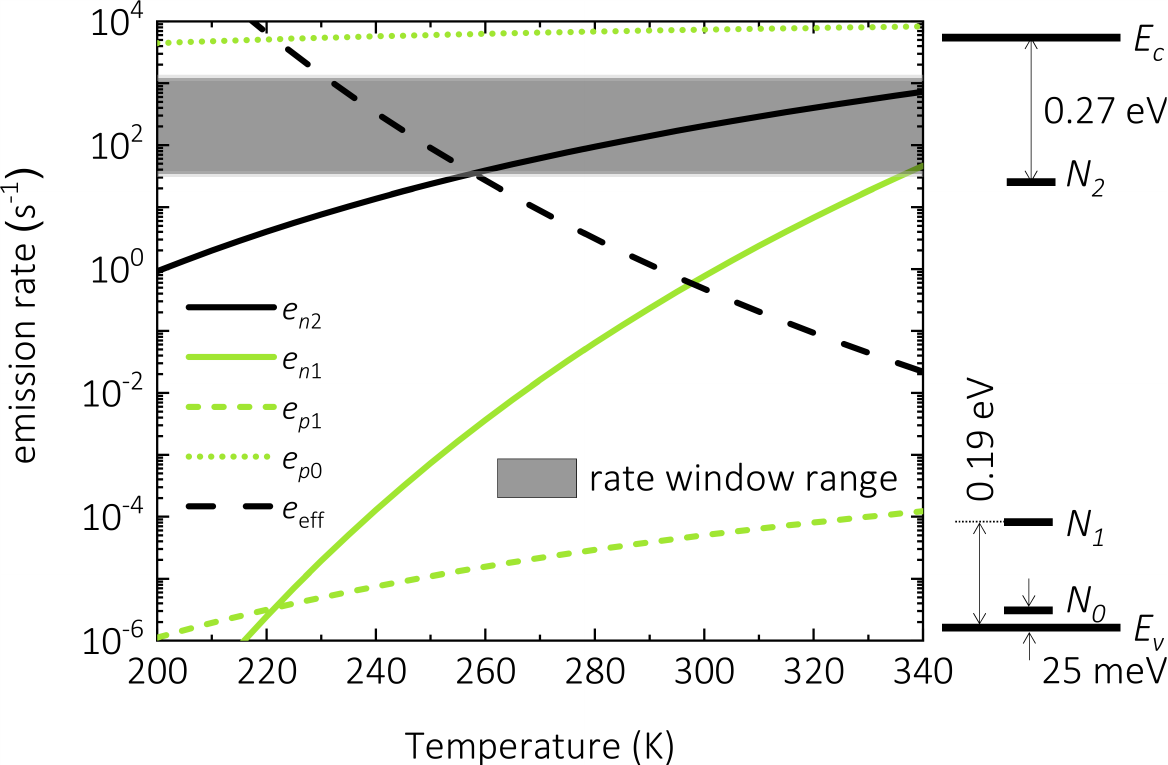}
\caption{\label{structure} Temperature variation of the emission rates given in Eq. (\ref{eN2}) and Eq. (\ref{rates}) using the estimated activation energies. The rates $e_{n2}$ and $e_{\textrm{eff}}$ appearing in the rate window range (gray zone) that are measured in the experiment are shown in black. The negative slope of $e_{\textrm{eff}}$ corresponds to an effective negative activation energy. The other rates, in green, fall outside the rate window range and are not directly measured in the experiment. To the right a schematic representation of the electronic structure of the Ga$_{\textrm{i}}$ center is shown.}
\end{figure}

An application of the boxcar method to the polarization-dependent PC transients measured over a temperature range of 200 K $< T <$ 340 K is shown in Fig. \ref{boxcar}. The boxcar signals, $\Delta I_{PC}$, correspond to a selection of rate windows ranging from $\{t_1,t_2\} = \{1\:\textrm{ms}, 3\:\textrm{ms}\}$ labeled (a) in Fig. \ref{boxcar}, to $\{36\:\textrm{ms}, 108\:\textrm{ms}\}$ labeled (r) in Fig. \ref{boxcar}. These two limiting cases define a rate window range shown in gray in Fig. \ref{structure}, with the maximum emission rate (a) equal to 1220 s$^{-1}$, and the minimum emission rate (r) equal to 34 s$^{-1}$. The full list of rate windows corresponding to the curves in Fig. \ref{boxcar} are given in the Supplementary Material. The signals obtained with a $\sigma$-polarized ($\pi$-polarized) filling pulse are shown in red (black). A four-component Gaussian fit to each curve (see Supplementary Material) reveals the possible presence of multiple, overlapping peaks, two of which are clear to the eye in Fig. \ref{boxcar}. The amplitude of the sharp, lower temperature peak is independent of filling pulse polarization, and its position shifts anomalously to higher temperatures as the rate window is moves to lower rates. The amlpitude of the broad, higher temperature peak \textit{does} depend on filling pulse polarization, with a lower amplitude measured for the $\sigma$-pump. Its position shifts to lower temperatures as the rate window is moved to lower rates as would be expected for a normal, thermally-activated process. This peak will be analyzed first.

\begin{figure}
\includegraphics[width=1\columnwidth]{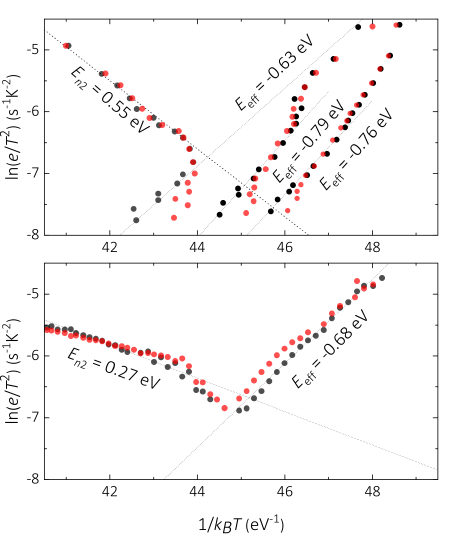}
\caption{\label{Arrhenius} Arrhenius plots generated from the boxcar (a) and inverse Laplace transform (b) analyses for $\pi$- (black) and $\sigma$- (red) polarized filling pulses. Activation energies obtained from the line slopes are shown.}
\end{figure}

The polarization-dependent capture processes occurring during the pol-PICTS filling pulse are shown (inset) in Fig. \ref{PLspectra}. For a $\pi$-pump the conduction band electron population changes on the timescales of the experimental rate window range (1 ms to 108 ms) due to charge re-emission from deep levels according to: \begin{equation} \label{conde} \frac{dn}{dt} = e_{n2}N_2 + e_{n1}N_1, \end{equation} where $e_{n2}$ and $e_{n1}$ are the electron emission rates from the $N_2$ and $N_1$ states respectively as shown in the insets of Fig. \ref{boxcar}. The valence band hole population changes according to: \begin{equation} \label{valh} \frac{dp}{dt} = e_{p1}N_1 + e_{p0}N_0, \end{equation} where $e_{p1}$ and $e_{p0}$ are the hole emission rates from the $N_1$ and $N_0$ states respectively, also shown inset in Fig. \ref{boxcar}. The overall PC transient like those shown in Fig. \ref{transients} results from a sum of Eq. (\ref{conde}) and Eq. (\ref{valh}), weighted for the transport coefficients.

In the case of the $\sigma$-pump, and in the limit of a 100 \% spin-polarized conduction electrons, electron capture to the $N_1$ is no longer allowed as explained above and indicated in Fig. \ref{PLspectra} (red inset). In this case there is no capture process which produces centers in the $N_2$ state so the first term in Eq. (\ref{conde}) is absent as shown schematically in Fig. \ref{boxcar} (red inset). The resulting reduction in the amplitude of the PC transient manifests itself as a reduction in the boxcar amplitude. Note again that if the conduction electron spin-polarization is lower than 100 \%, the boxcar amplitude is reduced (as observed) but is not zero. It is thus straightforward to associate the polarization-dependent boxcar peak with electron emission from the $N_2$ states at rate $e_{n2}$. Note that only the change in amplitude with filling pulse polarization is accounted for here. The absolute change in amplitude of both the red and black curves in Fig. \ref{boxcar} with a change in rate window arises because of transport effects, but does not affect the determination of activation energies \cite{balland1986}.

The temperature variation of each of the four Gaussian fit components to the boxcar signals in Fig. \ref{boxcar} are treated using the usual rate-window procedure to associate the temperature of each of them with an emission rate in order to produce an Arrhenius plot like that shown in Fig. \ref{Arrhenius}. One of the four peaks' does not change with the rate window and is not therefore associated with a thermally activated process -- its position on the Arrhenius plot is not shown.  In a thermally activated process of the form \begin{equation} \label{eN2} e_{n2}(T) = e_{n2}^0\exp\left[-E_{n2}/k_BT\right], \end{equation} the position of the Gaussian component will vary with rate window and yield a linear slope on the Arrhenius plot from which an activation energy of $E_{n2} =$ 0.55 eV is obtained. Note that this value is independent of the filling pulse polarization (black and red dots in Fig. \ref{Arrhenius}(a)). However, the significant peak overlap in the boxcar signal makes accurate estimation of activation energies difficult \cite{balland1986}. A better approach is the inverse Laplace transform \cite{evans2000} analysis which, when applied, does indeed yield a different, polarization-independent activation energy of 0.27 eV (see Fig. \ref{Arrhenius}(b)). This value is in agreement with a $n$-component exponential fit to the transients (not shown, see Supplementary Material), where $n$ is determined during the regularization part of the Laplace transform procedure, and is therefore considered to be a more realistic estimate of $E_{n2}$. The result is sketched in the electronic structure shown in Fig. \ref{structure} \cite{neumark1983}.

We now turn to the analysis of the lower temperature boxcar peak in Fig. \ref{boxcar} whose amplitude is independent of polarization, but which exhibits an anomalous shift to higher temperatures at longer rate windows. This peak is challenging to identify clearly on the boxcar signal, and two Gaussian components are using to obtained an acceptable fit (see Supplementary Material). These two peaks result in two lines with positive slope on the boxcar Arrhenius plot in Fig. \ref{Arrhenius}(a) with anomalous, negative activation energies ranging from -0.63 eV to at least -0.79 eV, again independent of filling pulse polarization. This large variation is again the result of the limitations of the boxcar method evoked above. The inverse Laplace transform analysis is much clearer, and shows the presence of a single emission process with an anomalous activation energy of -0.68 eV (see Fig. \ref{Arrhenius}(b)), as does the multi-exponential fit (not shown, see Supplementary Material). Since this approach is more reliable when boxcar peaks overlap, the effective activation energy is taken to be $E_{\textrm{eff}}=-0.68$ eV.

At first sight a negative activation energy is puzzling, but in fact is encountered in a number of interesting situations in chemistry. Examples include the oxidation of nitrous oxide \cite{mckee1995}, the cracking of n-paraffins \cite{wei1996}, and cell death rates as a result of hypothermia \cite{muench1996}. In each of these cases, the negative activation energy is explained by a multi-step process consisting of a fast, reversible reaction between the reactants and an intermediate product, which then proceeds via a slow, irreversible process to the final products. The essential idea is that the intermediate state must have a large activation energy so that it is preferentially emptied as temperature rises, thereby cutting off the route to the formation of the final products whose concentration will then \textit{decrease} with increasing temperature. An excellent generic description of this is given in Ref. \onlinecite{muench1996}.

Inspired by this, a similar configuration is proposed here, starting with a fast, reversible exchange between $N_1$ and $N_0$ states, followed by a slow conversion of $N_1$ states to $N_2$ states via hole emission according to:
\begin{equation} \label{reaction}
\ce{
N_0
<=>[\ce{e_{p0}}][\ce{e_{n1}}]
${\ce{N_1}}$
->[\ce{e_{p1}}]
${\ce{N_2}}$,
}
\end{equation} where, importantly, each of the emission rates appearing in the reaction and sketched in Fig. \ref{boxcar} (insets) are normally activated i.e. \begin{equation} \label{rates} \begin{aligned}
  e_{n1}(T) &= e_{n1}^0\exp\left[-E_{n1}/k_BT \right] \\
  e_{p0}(T) &= e_{p0}^0\exp\left[-E_{p0}/k_BT \right] \\
  e_{p1}(T) &= e_{p1}^0\exp\left[-E_{p1}/k_BT \right],
\end{aligned} \end{equation} with $E_{n1}$, $E_{p0}$, and $E_{p1}$ all \textit{positive}. To obtain an effective negative activation energy in the following, the condition \begin{equation} \label{condition} e_{p1} \ll e_{p0}, e_{n1} \end{equation} should be fulfilled \cite{muench1996}. Note that unlike the general assumption made throughout the literature that $N_0$ states are absent \cite{ivchenko2010}, in this picture the observation of a negative effective activation energy \textit{requires} the presence of $N_0$ states.

Ab-initio calculations of the electronic structure of Ga$_\textrm{i}$ in both GaAs \cite{baraff1985} and GaAsN \cite{laukkanen2012} suggest that the $N_0$ state lies close to the valence band edge. We therefore postulate that the activation energy for hole emission at rate $e_{p0}$ is within $k_BT\approx$ 25 meV of the valence band edge i.e. $E_{p0}=$ 0.025 eV. This is the \textit{only} energy in the electronic structure of the SDR-active center that is not determined from the pol-PICTS experiment. It suggests that $N_0$ is a shallow acceptor rather than a deep center. Its equilibrium density at room temperature is required to be small compared to the total Ga$_\textrm{i}$ density for large SDR \cite{ivchenko2010}, and this is ensured by taking the rate amplitude $e_{p0}^0$ in Eq. (\ref{rates}) to be sufficiently large compared to the other rate amplitudes. In the following a value of $e_{p0}^0 = 2 \times 10^4$ s$^{-1}$ is used which ensures that $e_{p0}$ is larger than all other emission rates (see green, dotted line in Fig. \ref{structure}). Two important observations should be immediately noted in Fig. \ref{structure}. Firstly, since $N_0$ is shallow, $e_{p0}$ only weakly depends on temperature, and secondly it falls outside the experimental rate window indicated in gray, meaning that $e_{p0}$ is too fast to be directly measured in the experiment.

Unlike $e_{p0}$, the emission rate $e_{n2}$ \textit{is} directly measured in the experiment as already discussed. Since its activation energy, $E_{n2} = 0.27$ eV, is already determined, a rate amplitude $e_{n2}^0 = 10^7$ s$^{-1}$ can be chosen so that $e_{n2}$ does fall in the rate window range above 260 K (solid black line in Fig. \ref{structure}) where it is observed experimentally in Fig. \ref{boxcar}. As the temperature drops $e_{n2}$ slows noticeably since $E_{n2}$ is relatively large, and approaches the bottom edge of the experimental rate window range around 260 K. Once it passes out of this range it is too slow to be experimentally measured. In this limit, the measurable rate of change of the photo-current transient due to electron emission into the conduction band in Eq. (\ref{conde}) becomes: \begin{equation} \label{freeze} \frac{dn}{dt} \approx e_{n1}N_1. \end{equation}

 Fig. \ref{structure} is obtained with rate amplitudes $e_{n1}^0 = 10^{15}$ s$^{-1}$ and $e_{p1}^0 = 10^{-1}$ s$^{-1}$. We emphasize that the values of the emission rate amplitudes are related to capture cross sections and can therefore vary over many orders of magnitude. However, their absolute values are not to be over-interpreted. Their relative values are however of interest. For example, the choice of $e_{n1}^0$ and $e_{p1}^0$ ensures the validity of Eq. (\ref{condition}) as seen by the relative positions of the three green lines in Fig. \ref{structure}, and the choice of $e_{n2}^0$ relative to $e_{n1}^0$, ensures that electron capture rates to $N_1$ are greater than those to $N_0$ i.e. $c_{n1} \gg c_{n0}$ which ensures large SDR as mentioned in the introduction. The choice of amplitudes also ensures that a steady-state between the $N_0$ and $N_1$ concentrations is established on timescales much shorter than that required for the final step in Eq. (\ref{reaction}). This steady state is expressed as $e_{p0}N_0 = e_{n1}N_1$ such that the the electron and hole contributions to the PC transient from Eq. (\ref{freeze}) and Eq. (\ref{valh}) can be written: \begin{equation} \label{electrons} \frac{dn}{dt} \approx e_{p0}N_0 \end{equation} and \begin{equation} \label{holes} \frac{dp}{dt} = \frac{e_{p0}e_{p1}}{e_{n1}}N_0 + e_{p0}N_0. \end{equation} 

Since $e_{p0}$ is outside the rate window range, the PC transient due to electron re-emission into the conduction band described by Eq. (\ref{electrons}) is immeasurably fast. This is also true for the second term in Eq. (\ref{holes}) which is part of the hole contribution to the PC transient. The anomalous peak in the boxcar signal occurring below 260 K in Fig. \ref{boxcar} is therefore attributed to the first term in Eq. (\ref{holes}). Using Eq. (\ref{rates}), $e_{\textrm{eff}} = e_{p0}e_{p1}/e_{n1}$ has an effective activation energy \begin{equation} \label{effective} E_\textrm{eff}=E_{p0}+E_{p1}-E_{n1} \end{equation} which is \textit{negative} if $E_{n1} > E_{p0}+E_{p1}$. With the measured, anomalous activation energy of $E_{\textrm{eff}} =$ -0.68 eV and $E_{p0} = $ 0.025 eV as previously stated, Eq. (\ref{effective}) yields $E_{p1}-E_{n1} = $ -0.705 eV. Combining this with knowledge of the gap from the PL spectrum in Fig. \ref{PLspectra}, $E_{p1}+E_{n1}$ = 1.09 eV, yields $E_{n1}$ = 0.9 eV and $E_{p1}$ = 0.19 eV. This result is sketched in the electronic structure in Fig. \ref{structure}. 

Fig. \ref{structure} shows the temperature dependence of $e_{\textrm{eff}}$ (black, dashed line) which exhibits the expected negative slope corresponding to a negative activation energy, and moreover, which crosses the rate window range in the temperature range 220 K $< T <$ 260 K where the anomalous peak in Fig. \ref{boxcar} appears. The figure is extremely useful to help physically summarize the observation of a negative activation energy. Consider a temperature rise from 200 K. On the scale of the changes in the emission rates with temperature, $e_{p0}$ is essentially constant because of its small activation energy. The generation of centres in the $N_1$ state via this process is therefore independent of temperature. On the other hand, electron emission at rate $e_{n1}$ increases significantly with temperature because of its large activation energy. Consequently, the rapid steady-state between $N_0$ and $N_1$ shifts towards an increase (decrease) in the density of $N_0$ ($N_1$) states. Note also that the re-population of $N_1$ states via electron emission from the $N_2$ states cannot compensate for this since $e_{n2}$ is less sensitive to temperature than $e_{n1}$. Increasing temperature therefore \textit{depopulates} the $N_1$ states.

Return now to the first term on the right hand side in Eq. (\ref{valh}). If the depopulation of $N_1$ with increasing temperature is faster than the thermally activated increase in $e_{p1}$, then this term becomes smaller with increasing temperature. Since it is the only term in the rate window range below 260 K, the measured emission rate in the PC transient proportional to $dp⁄dt$ will decrease anomalously with temperature resulting in an apparent negative activation energy. To paraphrase from Ref. \onlinecite{wei1996} ``This phenomenon is contrary to normal expectations, and is a consequence of the competition between two effects: the increase of intrinsic kinetics with temperature (here $e_{p1}$), and the decrease of \ldots the concentration of active intermediates with temperature (here $N_1$).''

In conclusion, using a novel pol-PICTS approach that adds spin-sensitivity to the usual PICTS method, the electronic structure of the paramagnetic center responsible for the spectacular SDR observed in dilute nitrides has been estimated for the first time. The result, shown sketched in Fig. \ref{structure}, is as important as the crystallographic identification of the Ga$_{\textrm{i}}$ center \cite{wang2009} in that the electronic structure fundamentally determines the nature of the electronic states as donors, acceptors, traps, or recombination centers. The observations are consistent with the presence of a shallow, acceptor-like state ($N_0$) which has hitherto been ignored in coupled spin/charge models. Since these models only approximately reproduce the SDR properties of these alloys, the result can inform improvements to them. The state energies are only in approximate agreement with ab initio calculations of the electronic structure that assume particular arrangements of nitrogen atoms around the Ga$_{\textrm{i}}$ interstitial \cite{laukkanen2012}. This should inspire new attempts to identify the local chemical environment of the SDR-active interstitial.

\begin{acknowledgments}
ACU acknowledges support of the FASIC program for travel support (\textit{partenariat Hubert Curien franco-australien}). The authors thank N. Vast and Y. Cho for useful discussions.
\end{acknowledgments}

\bibliographystyle{apsrev4-2}
\bibliography{References}

\end{document}